\documentclass[twocolumn,superscriptaddress,aps,prx]{revtex4-1}
\pdfoutput=1
\usepackage[english]{babel}
\usepackage[utf8]{inputenc}

\usepackage{microtype}
\usepackage{mathtools}
\usepackage{graphicx}
\usepackage[hidelinks]{hyperref}

\usepackage{csquotes}
\usepackage{standalone}
\usepackage{siunitx}
\usepackage{pgfplots}
\usepackage{subcaption}

\graphicspath{{./}{images/}}

\makeatletter
\colorlet{maroon}{red!80!black}
\usetikzlibrary{
	shapes,
	positioning,
	arrows,
	decorations.markings,
	calc,
	fit,
	patterns,
}
\tikzstyle{every picture}+=[font=\sffamily]

\tikzstyle{laserstartarrow} = [
	draw=maroon,
	line cap=round,
	decoration={markings,mark=at position 0.1 with{\arrow[maroon]{latex'}}},
	postaction={decorate},
	mark options={color=maroon}
]

\tikzstyle{laser} = [
	draw=maroon,
	line cap=round,
]
\tikzstyle{laserarrow} = [
	line cap=round,
	draw=maroon,
	decoration={markings,mark=at position 0.56 with{\arrow[maroon]{latex'}}},
	postaction={decorate},
	mark options={color=maroon}
]
\tikzstyle{laserdetourarrow} = [
	line cap=round,
	draw=maroon,
	decoration={markings,mark=at position 0.55 with{\arrow[rotate=-20,maroon]{latex'}}},
	postaction={decorate},
	mark options={color=maroon}
]

\tikzstyle{right-hand-mirror} = [
	draw,
	postaction=decorate,
	decoration={
		markings,
		mark=between positions 0.015 and 0.98 step 0.1072
		with {\draw(0,0)-- (120:3pt);}
	}
]

\tikzstyle{left-hand-mirror} = [
	draw,
	postaction=decorate,
	decoration={
		markings,
		mark=between positions 0.015 and 0.98 step 0.1072
		with {\draw(0,0)-- (60:3pt);}
	}
]

\tikzstyle{arrow} = [
	draw,
	thick,
	-latex'
]

\tikzstyle{detector} = [
	draw,
	thick
]

\tikzstyle{beamsplitter} = [
	fill=gray,
	fill opacity=0.3,
]
\tikzstyle{phasemodulator} = [
	fill=yellow,
	fill opacity=0.5,
]
\tikzstyle{coupler} = [
	draw=blue!80!white,
	thick,
	draw opacity=0.7,
]

\begin{document}

\title{Hacking energy-time
entanglement-based systems with classical light}

\author{Jonathan Jogenfors} \affiliation{Institutionen för Systemteknik,
  Linköpings Universitet, 581 83 Linköping, Sweden.}%
\author{Ashraf M. Elhassan} \affiliation{Physics Department, Stockholm
  University, 106 91, Stockholm, Sweden.}%
\author{Johan Ahrens} \affiliation{Physics Department, Stockholm University,
  106 91, Stockholm, Sweden.}%
\author{Mohamed Bourennane} \affiliation{Physics Department, Stockholm
  University, 106 91, Stockholm, Sweden.}%
\author{Jan-Åke Larsson} \email{jan-ake.larsson@liu.se}%
\affiliation{Institutionen för Systemteknik,
  Linköpings Universitet, 581 83 Linköping, Sweden.}

\begin{abstract}
	Photonic systems based on energy-time
	entanglement have been proposed to test local realism using the Bell
	inequality. A violation of this inequality normally also certifies security
	of device-independent quantum key distribution, so that an attacker cannot
	eavesdrop or control the system. Here, we show how this security test can be
	circumvented in energy-time entangled systems when using standard avalanche
	photodetectors, allowing an attacker to compromise the system without
	leaving a trace. With tailored pulses of classical light we reach Bell
	values up to 3.63 at 97.6\% detector efficiency which is an extreme
	violation. This is the first demonstration of a violation-faking source that
	both gives tunable violation and high detector efficiency. The implications
	are severe: the standard Clauser-Horne-Shimony-Holt inequality cannot be
	used to show device-independent security for standard postselecting
	energy-time entanglement setups. We conclude with suggestions of improved
	tests and experimental setups that can re-establish device-independent
	security.
\end{abstract}

\maketitle

\let\thefootnote=\relax \footnotetext{$^*$These authors contributed equally to
  this paper. $^\dag$Institutionen för Systemteknik, Linköpings
  Universitet, 581 83 Linköping, Sweden.  $^\ddag$Physics Department,
  Stockholm University, 106 91, Stockholm, Sweden. $^\S$Correspondence should
  be addressed to J.-Å.\ L.\ (email:
  \href{mailto:jan-ake.larsson@liu.se}{jan-ake.larsson@liu.se})}

\noindent
A Bell experiment~\cite{Bell1964} is a bipartite experiment that can be used
to test for pre-existing properties that are independent of the measurement
choice at each site. Formally speaking, the experiment tests if there is a
\enquote{local realist} description of the experiment, that contains these
pre-existing properties. Such a test can be used as the basis for security of
Quantum Key Distribution~\cite{Bennett1984,Ekert1991} (QKD). QKD uses a
bipartite quantum system shared between two parties (Alice and Bob), that
allows them to secretly share a cryptographic key. The first QKD protocol
\cite{Bennett1984} (BB84) is based on quantum uncertainty
\cite{Heisenberg1927} between non-commuting measurements, usually of photon
polarization. The Ekert protocol~\cite{Ekert1991} (E91) bases security on a
Bell test instead of the uncertainty relation. Such a test indicates, through
violation of the corresponding Bell inequality, a secure key distribution
system. This requires quantum entanglement, and because of this E91 is
also called entanglement-based QKD.

To properly show that an E91 cryptographic system is secure, or alternatively,
that no local realist description exists of an experiment, a proper violation
of the associated Bell inequality is needed. As soon as a proper violation is
achieved, the inner workings of the system is not important anymore, a fact
known as device-independent security~\cite{Acin2006,Acin2007}, or a
loophole-free test of local realism~\cite{Larsson2014}. In the security
context, the size of the violation is related to the amount of key that can be
securely extracted from the system.  However, a \emph{proper} (loophole-free)
violation is difficult to achieve. For long-distance experiments, photons is
the system of choice and one particularly difficult problem is to detect
enough of the photon pairs; this is known as the efficiency loophole
\cite{Pearle1970,Garg1987, Larsson1998}.

If the violation is not good enough, there may be a local realist description
of the experiment, giving an insecure QKD system. Even worse, an attacker
could control the QKD system in this case. One particular example of this
occurs when using avalanche photodetectors (APD:s) which are the most commonly
used detectors in commercial QKD systems: these detectors can be controlled by
a process called \enquote{blinding}~\cite{Lydersen2010} which enables control
via classical light pulses. When using photon polarization in the system, and
if the efficiency is low enough in the Bell test, the quantum-mechanical
prediction can be faked in such a controlled system~\cite{Larsson2002a,Gerhardt2011}. This means that the (apparent) Bell
inequality violation can be faked, making a QKD system seem secure while it is
not. Note that a proper (loophole-free) violation cannot be faked in this
manner.

In this paper we investigate energy-time entangle\-ment-based systems in
general and the Franson interferometer~\cite{Franson1989} in particular.
Traditional polarization coding is sensitive to polarization effects caused by
optical fibers~\cite{Gisin2002} whereas energy-time entanglement is more robust
against this type of disturbance. This property has led to an increased
attention to systems based on energy-time entanglement since it allows a design
without moving mechanical parts which reduces complexity in practical
implementations. A number of applications of energy-time entanglement, such as
the QKD, quantum teleportation and quantum repeaters are described
in~\cite{Gisin2007}.

It is already known that a proper Bell test is more demanding to achieve in
energy-time-entanglement systems with postselection~\cite{Aerts1999,Jogenfors2014}, but also that certain assumptions
on the properties of photons reduce the demands to the same level as for a
photon-polarization-based test~\cite{Franson1999,Franson2009}. The property in
question is particle-like behavior of the photon: that it does not
\enquote{jump} from one arm of an interferometer to the other. Now, clearly,
classical light pulses cannot \enquote{jump} from one arm to the other, so the
question arises: is it at all possible to control the output of the detectors
using classical light pulses, to make them fake the quantum correlations?
Below, we answer this question in the positive, give the details of such an
attack, and its experimental implementation.  Moreover, not only are faked
quantum correlations possible to reach at a faked detector efficiency of 100\%,
but even the extreme predictions of nonlocal Popescu-Rohrlich boxes
\cite{Popescu1994} are possible to fake at this high detector efficiency. Such
extreme predictions reaches the algebraic maximum 4 of the CHSH inequality, and
would make a QKD system user suspicious; an attacker would of course not exceed
the quantum bound $2\sqrt2$~\cite{Cirelson1980}.

\begin{figure}[t]
%	\centering
	% \begin{subfigure}{\linewidth}
	% 	\includestandalone{fransoninterferometer}
	% 	\caption{Diagram}
	% \end{subfigure}
%	\begin{subfigure}{0.8\linewidth}
		\includegraphics[width=\linewidth]{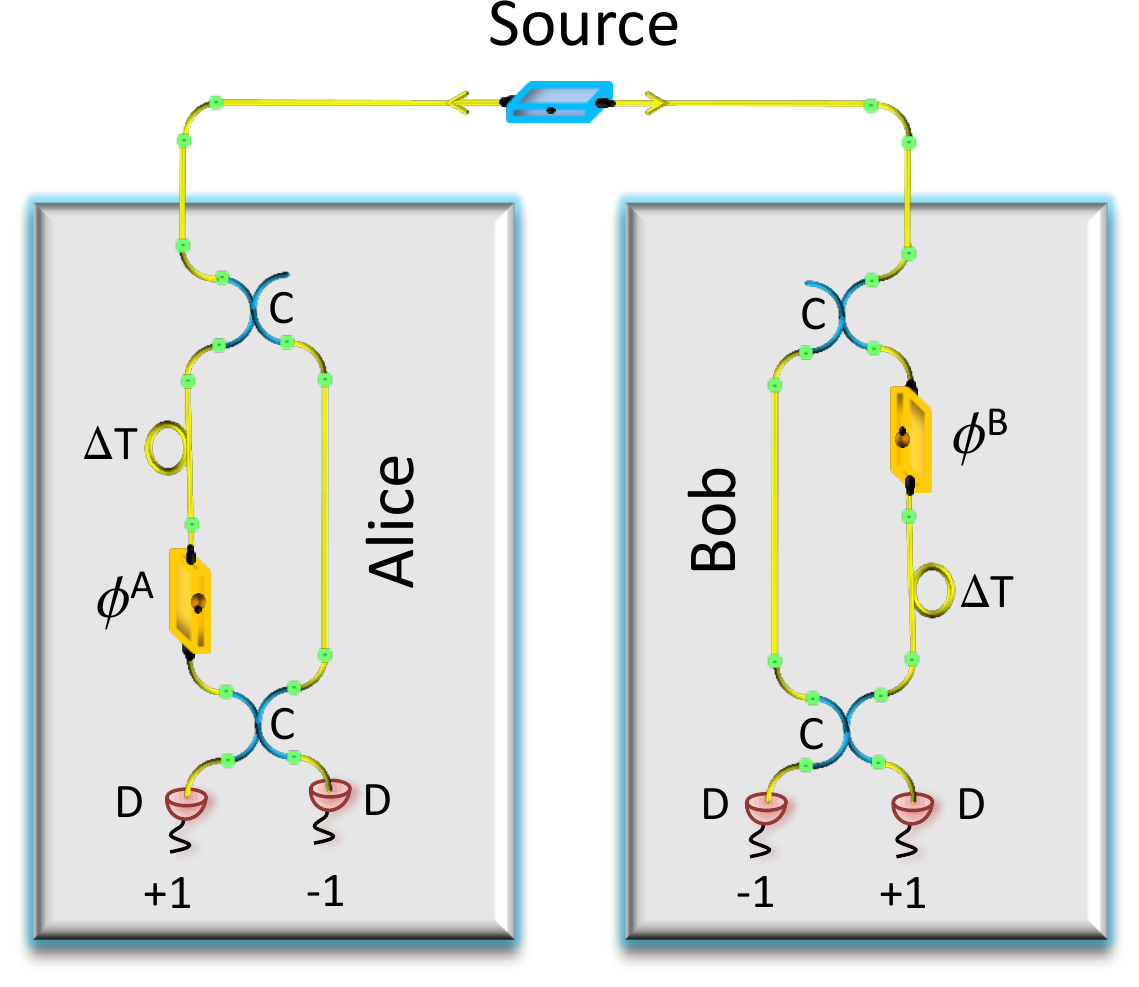}
		\caption{Experimental setup of the Franson interferometer. The
			setup consists of a source,
			$2\times2$-couplers (C), delay loops ($\Delta T$),
		phase modulators $\phi^A$ and $\phi^B$ and detectors (D).}
%	\end{subfigure}
%	\caption{The Franson interferometer}
	\label{fig:franson}
\end{figure}

\section{Bell's inequality and the Franson interferometer}
A Bell test of device-independent security, alternatively local realism, is
always associated with a Bell inequality.  The relevant part of the E91 QKD
protocol up to and including the Bell test looks as follows. The general setup
is a central source connected to two measurement sites, one at Alice and the
other at Bob.  The source prepares an entangled quantum state and distribute
it to Alice and Bob who each can choose between a number of measurement
settings for their devices. The output can take the values $-1$, 0, or $+1$,
denoting for example horizontal polarization, non-detection, and vertical
polarization. Here we are considering a pulsed source so that there are
well-defined experimental runs, and therefore also well-defined non-detection
events. Alice selects a random integer $j\in\{1,2,3\}$ and performs the
corresponding measurement $A_j$. Bob does the same with a random number
$k\in\{2,3,4\}$ and measurement $B_k$. The quantum state and measurements are
such that if $j=k$, the outcomes are highly (anti-)correlated. This
preparation and measurement process is performed over and over again until
enough data has been gathered.

After a measurement batch has been completed, Alice and Bob publicly announce
which settings $j$ and $k$ were used (but not the corresponding
outcomes!). They can then determine which measurements used the same settings
$j=k$ and use the highly (anti-)correlated outcomes for key generation. The
remaining outcomes corresponding to $j\neq k$ can be used for security
testing, in the Bell~\cite{Bell1964}-CHSH~\cite{Clauser1969} inequality
\begin{equation}
	\begin{split}
		S_2=&\left|E(A_1B_2)+E(A_3B_2) \right|\\
		&+\left|E(A_3B_4)-E(A_1B_4) \right|\le 2, \label{eqn:CHSH}
	\end{split}
\end{equation}
where $E(A_jB_k)$ is the expected value of the product, often called
\enquote{correlation} in this context. If the experimental $S_2$ is larger
than 2 there is a violation, and the system is secure; there can be no local
realist description of the experiment. The size of the violation is related to
output key rate; the maximal quantum prediction is $2\sqrt2$.

However, a proper violation is difficult to achieve. There are a number ways
that the test can give $S_2>2$ but still fail, a number of loopholes
\cite{Larsson2014}. The most serious one here is the detector efficiency
loophole, that non-detections or zeros are not properly taken into account. If
the zeros are ignored, conditioning on detection at both sites gives the
conditional correlation $E(A_jB_k|$coinc.$)$, and a modified bound
\cite{Garg1987,Larsson1998}
\begin{equation}
	\begin{split}
		S_{2,\text{c}}=&\left|E(A_1B_2|\text{coinc.})+E(A_3B_2|\text{coinc.}) \right|\\
		&+\left|E(A_3B_4|\text{coinc.})-E(A_1B_4|\text{coinc.}) \right|
		\le \tfrac4\eta-2. \label{eqn:CHSH_eff}
	\end{split}
\end{equation}
The efficiency $\eta$ is the ratio of coincidences to local detections
\cite{Larsson1998}, and needs to be above 82\% for the quantum value to give a
violation. This is ignored in current experiments, almost
\cite{Giustina2013,Christensen2013,Larsson2014a} without exception. In the
context of QKD, ignoring the zeros is allowed only if the attacker (Eve)
cannot control the detectors to make no-detections depend on the local
settings $j$ and $k$. Unfortunately, the commonly used APD:s can be controlled
\cite{Lydersen2010,Gerhardt2011} unless extra precautions are taken.

For this paper we have investigated a quantum device based on
energy-time entanglement with postselection. While the results presented below
are acquired from this particular device, the results apply to any such system.
The Franson
interferometer~\cite{Franson1989} is shown in Fig.\nobreakspace \ref
{fig:franson} and is built around a source emitting time-correlated
photons to both Alice and Bob. The unbalanced Mach-Zehnder interferometers
have a time difference $\Delta T$ between the paths.  In our pulsed setting,
the time difference between a late and early source emission is $\Delta T$,
giving rise to interference between the cases \enquote{early source emission,
	photons take the long path} and \enquote{late source emission, photons take
	the short path}. There will be no interference if the photons \enquote{take
	different paths} through the analysis stations, and those events are
discarded as non-coincident in a later step.

\begin{figure*}[t]
	\centering
	\begin{subfigure}{0.4\linewidth}
		\includegraphics{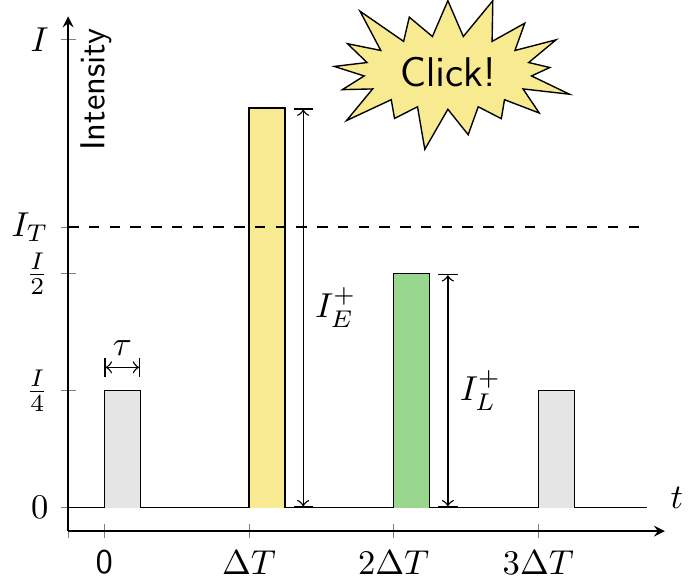}
		\caption{Constructive interference at the $+$ output gives a large early
      timeslot intensity and a corresponding click.\label{fig:click}}
	\end{subfigure}\quad
	\begin{subfigure}{0.4\linewidth}
		\includegraphics{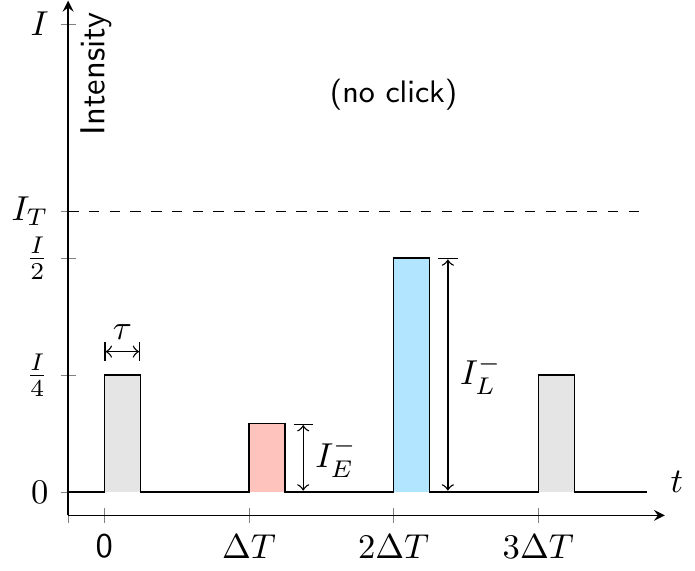}
		\caption{Destructive interference at the $-$ output gives a small early
      timeslot intensity and no corresponding click.\label{fig:noclick}}
	\end{subfigure}
	\caption{The blinding attack causes the detector to click only for pulses of
    greater intensity than $I_T$. If Eve sends three pulses of equal intensity
    $I$, they will arrive as four after the interferometer. By changing the
    phase shifts $\omega_E$ and $\omega_L$ between the pulses at the source,
    she can control the intensity of the early and late middle pulses at the
    $\pm$ output ports, giving clicks as desired. Here, $\phi=0$,
    $\omega_E=\pi/8$, and $\omega_L=\pi/4$. The first and last pulse have a
    constant intensity of $I/4$. \label{fig:intensity}}
\end{figure*}

The analysis stations have variable phase modulators and the setting choices
are $\phi^A_j$ for measuring $A_j$ at Alice and $\phi^B_k$ for measuring $B_k$
at Bob. The quantum state is such that, given coincident detection, the
correlation between $A_j$ and $B_k$ is high if $\phi^A_j+\phi^B_k=0$. In the
absence of noise, the correlation between Alice's and Bob's outcomes will
be~\cite{Franson1989}
\begin{equation}
	E(A_jB_k|\text{coinc.})=\cos(\phi^A_j+\phi^B_k)
	\label{eqn:franson-correlation}
\end{equation}
This again violates the CHSH inequality (\ref{eqn:CHSH}), but only if the
postselection is ignored~\cite{Aerts1999}. When the postselection is taken
into account one arrives at the inequality (\ref{eqn:CHSH_eff}) with
$\eta=50\%$, giving a bound of 6 which is no restriction. The question is
rather if the system can be controlled by Eve, to fake the violation.

\section{Faking the Bell inequality violation}\label{sec:faking}

An eavesdropper (Eve) performs the attack by replacing the source with a
faked-state generator that blinds the APD:s (see Fig.\nobreakspace \ref
{fig:intensity}) and makes them click at chosen instants in time.  The blinding
is accomplished using classical light pulses superimposed over continuous-wave
(CW) illumination~\cite{Lydersen2010}. In normal operation, an APD reacts to
even a single incoming photon.  A photon that enters the detector will create an
avalanche of electrical current which results in a signal, or \enquote{click},
when the current crosses a certain threshold.  The avalanche current is then
quenched by lowering the APD bias voltage to below the breakdown voltage, making
the detector ready for another photon, and resulting in so-called Geiger mode
operation.  Under the influence of continuous-wave (CW) illumination, the
quenching circuitry will make the current through the APD:s proportional to the
power of the incoming light. This will change the behaviour of the APD into
so-called linear mode, more similar to a classical photodiode.  It will no
longer react to single photons, nor register clicks in the usual Geiger-like way
and is therefore said to be \enquote{blind}.  Appropriate choice of CW
illumination intensity will make the APD insensitive to single photons yet still
register a click when a bright pulse of classical light is superimposed over the
CW illumination~\cite{Lydersen2010}.

What remains is to construct classical light pulses that
will give clicks in the way that Eve desires, violating the Bell inequality test
for the Franson interferometer.  Eve uses pulses with intensity $I$ and pulse
length $\tau\ll\Delta T$ intermingled with the CW light that blinds the APD:s. A
single pulse emitted by the source will be split when traveling through the
interferometer, resulting in two pulses in each output port with intensity $I/4$
each.  If instead two pulses are emitted, separated by $\Delta T$ and with phase
difference $\omega$, these two pulses will split to three. The middle pulse of
the three is built up by two parts, so that the $\pm1$ outputs show
interference,
\begin{equation}\label{eqn:I2}
	\begin{split}
		I^+(\phi,\omega)&= I\cos^2\big(\tfrac{\phi+\omega}2 \big)\\
		I^-(\phi,\omega)&= I\sin^2\big(\tfrac{\phi+\omega}2 \big),\\
  \end{split}
\end{equation}
where $\phi$ is the phase setting of the local analysis station. The chosen
$\omega$ controls the $\phi$ dependence of the output. For example, if $I$ is
just less than $2I_T$ and $\omega=0$, there will be a $+1$ click for
$|\phi|<\pi/2$ and a $-1$ click otherwise.

\begin{figure*}[t]
	\centering
	\includegraphics{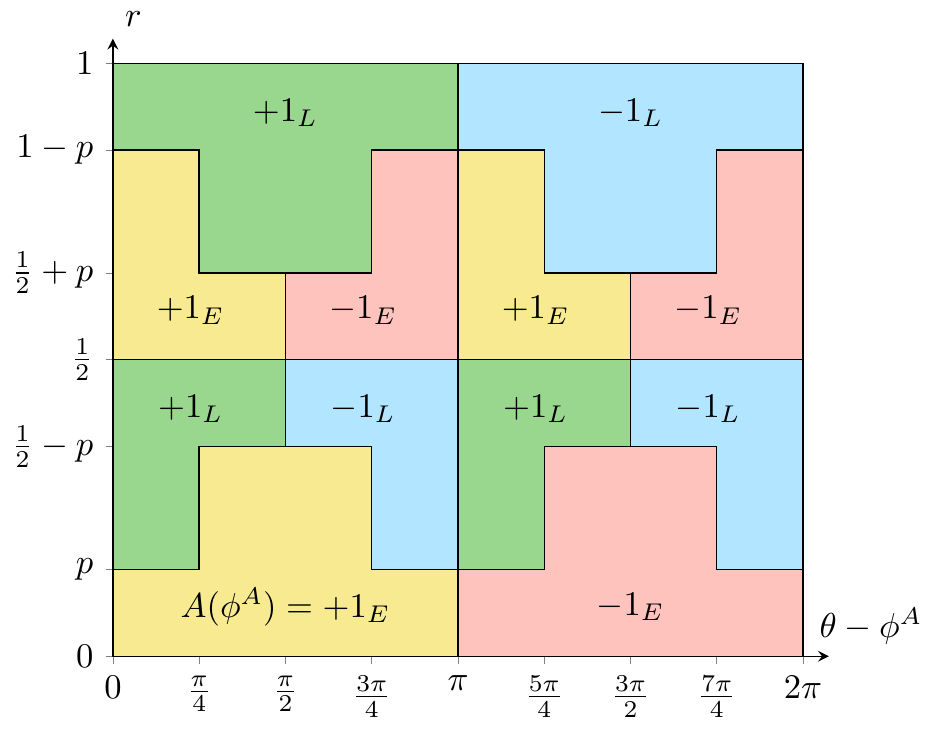}%
	\includegraphics{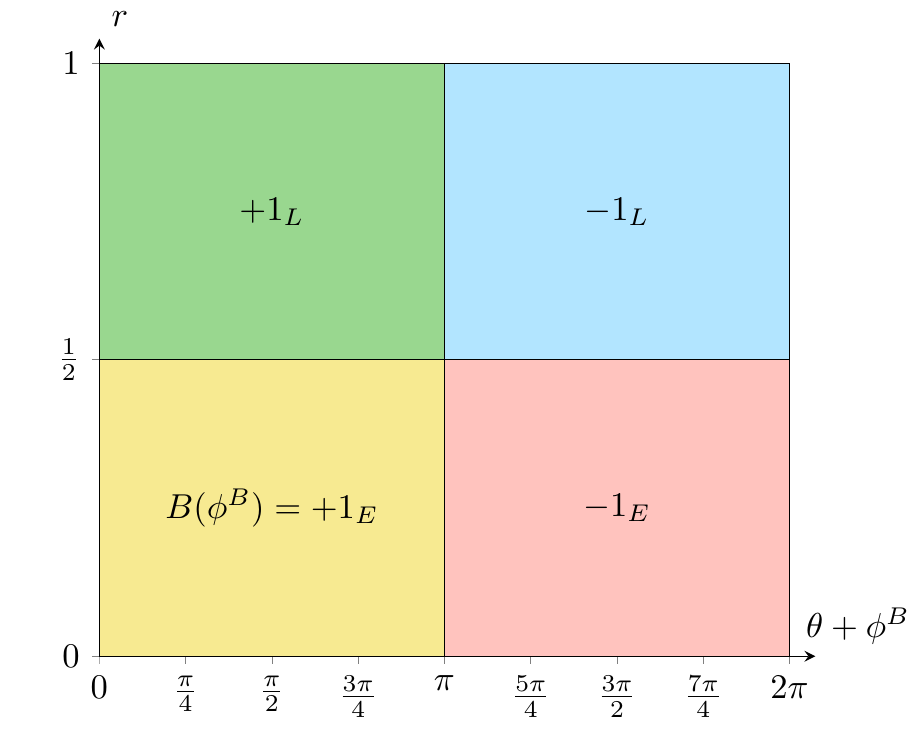}
	\caption{Discretized local hidden-variable model~\cite{Aerts1999}, that can
    give any Bell value between 2 and 4.  The hidden variables are $0\le r<1$
    (a real number in the unit interval) and and $\Theta=n\frac\pi4$ where
    $0\leq n \leq 7$ is an integer. The value $0\le p\le1/4$ can be chosen
    freely, and the output Bell value is $S_2=4-8p$, so that the ``classical''
    $S_2=2$ is obtained with $p=1/4$, the ``quantum'' $S_2=2\sqrt2$ is
    obtained with $p=(2-\sqrt2)/4$ (as in the figure), and the
    ``nonlocal-box'' $S_2=4$ is obtained with $p=0$, all at 100\% efficiency
    and 50\% postselection. }
	\label{fig:lhv}
\end{figure*}

However, this is not enough to fake the Bell violation, because the detection
time needs to depend on the local setting~\cite{Aerts1999}. To enable this,
Eve makes the source emit a group of three pulses separated by $\Delta T$,
with phase difference $\omega_E$ between the first and second pulse, and
$\omega_L$ between the second and third pulse. When this pulse train passes
through the interferometer, the output is four pulses, where the two center
pulses have controllable intensity because of interference. The intensities
for these two (Early/Late) pulses are
\begin{equation}\label{eqn:I3}
	\begin{split}
		I_E^+(\phi,\omega_E)&
    =I\cos^2\big(\tfrac{\phi+\omega_E}2\big)\\
    I_E^-(\phi,\omega_E)&
    =I\sin^2\big(\tfrac{\phi+\omega_E}2\big)\\
    I_L^+(\phi,\omega_L)&
    =I\cos^2\big(\tfrac{\phi+\omega_L}2\big)\\
    I_L^-(\phi,\omega_L)&
    =I\sin^2\big(\tfrac{\phi+\omega_L}2\big).
	\end{split}
\end{equation}
For example, with the same choice of $I$ as above, $\omega_E=0$, and
$\omega_L=\pi/2$, there will be an early $+1$ click if $\phi=0$, and a late
$-1$ click if $\phi=\pi/2$.  Note that the pulse trains to Alice and Bob can
be chosen independently.

To fake the violation of the Bell inequality, Eve uses the local hidden
variable (LHV) model in Fig.\nobreakspace \ref {fig:lhv}, which is a discretized version of an
earlier known model~\cite{Aerts1999}. This technique can be extended to give
the entire set of quantum predictions but we have here chosen to focus on the
settings used for the present Bell test: $\phi^A_1=0$, $\phi^A_3=\pi/2$,
$\phi^B_2=-\pi/4$, and $\phi^B_4=-3\pi/4$ so that only $\theta$ in increments
of $\pi/4$. Eve randomly selects the hidden variables $r$ and $\theta$, and
reads off the desired results for the two settings at Alice. If the results
are in the same time slot, she uses two pulses, and can directly calculate the
needed phase difference. If the results are in different time slots (this only
happens for Alice), Eve uses three pulses and calculates the two phase
differences.  The same $r$ and $\theta$ are used to calculate the phase
difference for Bob. Repeating this procedure will produce random outcomes (to
Alice and Bob) that give exactly the quantum predictions for the mentioned
settings, violating the Bell-CHSH inequality.

By adjusting parameters of the LHV model we can go even further and produce
Bell values up to and including the value 4, see Fig.\nobreakspace \ref {fig:lhv}. But remember
that Alice and Bob would be very confused if their security test displayed the
value 4 as that would mean that their experiment consists of unphysical and
nonlocal Popescu-Rohrlich (PR) boxes~\cite{Popescu1994}.  Eve would avoid this
but could in principle use this possibility to negate the effects of noise, by
adjusting the LHV model to exceed the quantum prediction, knowing that the
noise will lower the violation to below the quantum bound again.

\section{Experimental results}

\begin{figure*}[t]
	\centering
	\includegraphics[width=.8\linewidth]{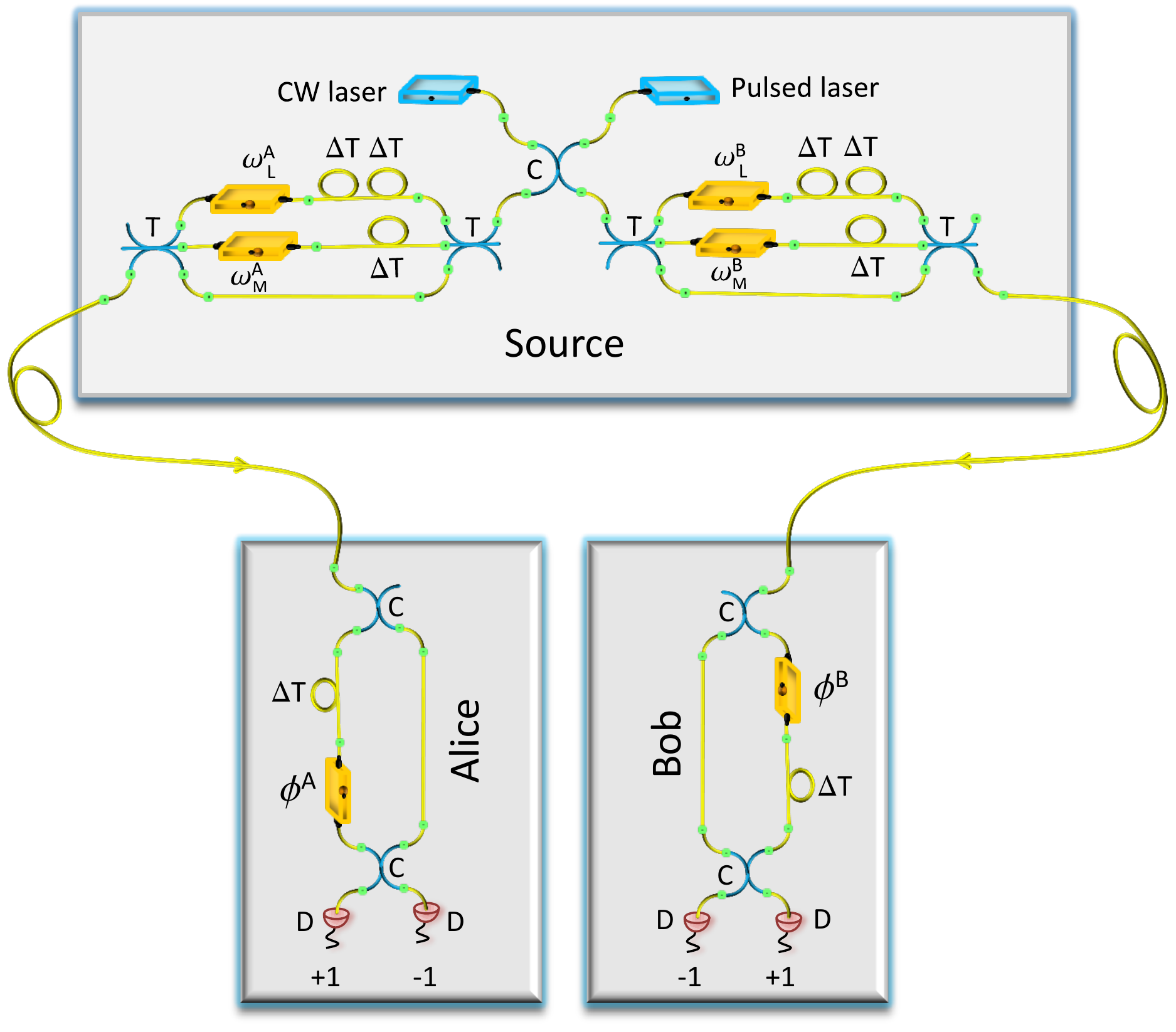}
	\caption{Experimental setup of the attack on the Franson interferometer. The
		setup consists of a continuous-wave (CW) laser for blinding the
		detectors, a pulsed laser for generating the bright classical light
		pulses, fiber-optical couplers (C) delay loops
		($\Delta T$), phase modulators ($\omega$ and $\phi$) and detectors (D).}
	\label{fig:experiment}
\end{figure*}

The attack was experimentally implemented as shown in Fig.\nobreakspace \ref
{fig:experiment}.  Built using standard fiber-optical components, it is designed
to meet the requirements set in the section \nobreakspace \ref{sec:faking}.
The continuous wave is produced by a CW laser while the pulses are created by a
pulsed laser.  These two light sources are combined at a fiber-optic $2\times2$
coupler and then split into one beam for Alice and one for Bob. Each of these
beams are then sent into a fiber-optic $3\times3$ coupler (tritters) that
equally divides them into three arms. The first arm consists of a $\Delta T$
delay loop and a phase modulator $\omega_E$, the second arm has two $\Delta T$
delay loops and a phase modulator $\omega_M$ (so that
$\omega_L=\omega_M-\omega_E$) while the third arm performs no action. The three
arms are then combined by a second $3\times3$ coupler into one output port that
creates the output of the faked state source generator.

The source sends bright light pulses with the setting and phase difference(s)
to Alice's and Bob's analysis stations in the
Franson interferometer. Each of the two analysis stations are constructed in a
similar fashion: Two fiber-optic $2\times2$ couplers and one delay loop $\Delta
T$ and a phase modulator $\phi^A$ (Alice's side) or $\phi^B$ (Bob's side). Gated
APD:s were used as detectors. This type of detector reduces the dark counts by
raising the bias voltage above the breakdown voltage only for a short time
period when an incoming photon is expected. These gated APD:s are still
vulnerable to the blinding attack described above even if the details of the
attack are slightly different. Since the CW power becomes unevenly distributed
between detectors, the efficiency of the blinding was affected. This imbalance
was avoided by installing digital variable attenuators at the output ports. In
addition, optical isolators were placed in front of the detectors in order to
prevent crosstalk.

\begin{figure}[t]
	\centering
	\includegraphics{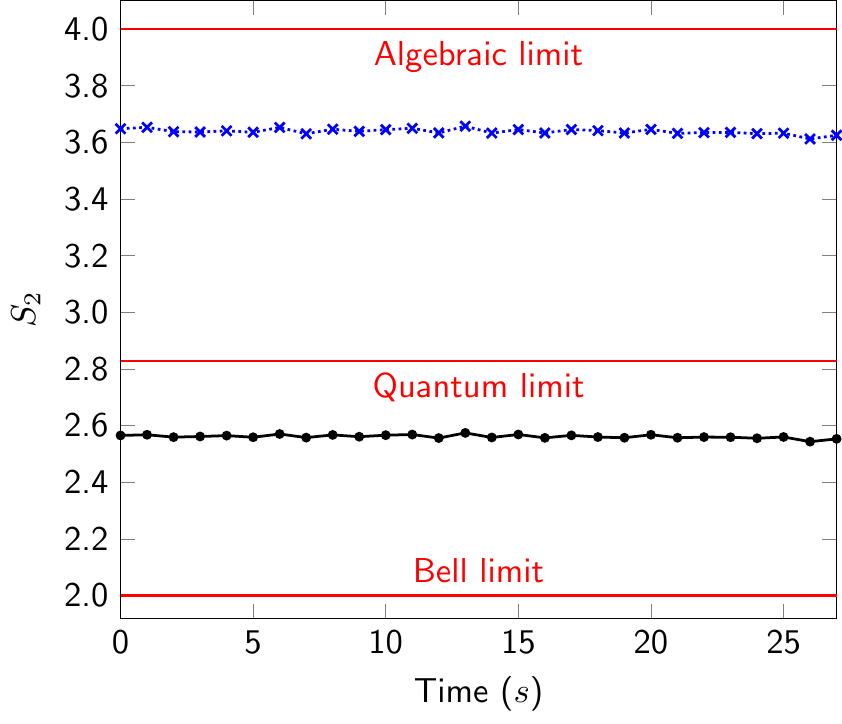}
	\caption{The faked Bell of our source is $2.5615\pm 0.0064$ (solid black
    line) which clearly violates the CHSH inequality $S_2\leq2$. It is
    possible to increase the faked Bell value up to $3.6386 \pm 0.0096$
    (dotted blue line, data for timeslots where $p\le r<1/2-p$ or $1/2+p\le
    r<1-p$). In both cases the efficiency is \SI{97.6}{\percent}. Each point
    in the diagram corresponds to the $S_2$ value for \SI{1}{\second} worth of
    data.\label{fig:bellplot}}
\end{figure}

Joint Alice-Bob trials were performed with the pulse amplitudes as described by
eqs.\nobreakspace \textup {(\ref {eqn:I2})} and\nobreakspace  \textup {(\ref
{eqn:I3})} and depicted in Fig.\nobreakspace \ref {fig:intensity}.  At the
desired detector and timeslot a \enquote{click} will be forced
(Fig.\nobreakspace \ref {fig:click}) by constructive interference while
destructive interference causes \enquote{no click} (Fig.\nobreakspace \ref
{fig:noclick}).  The sampling time used was \SI{1}{\second} and each experiment
was run for at least \SI{27}{\second} (see Fig.\nobreakspace
\ref{fig:bellplot}). At each point in time, the joint probabilities of Alice's
and Bob's outcomes are computed from the detector counts and these were then
used to determine the Bell value.  Note that the early and late timeslots are
measured in different experimental runs.  The average faked Bell value is
\begin{equation}
	S_2=2.5615\pm 0.0064
  \label{eq:1}
\end{equation}
which clearly violates the Bell bound 2.  Our source has a repetition rate of
of \SI{5}{\kilo\hertz}, and the average rate of clicks is
\SI{4.88}{\kilo\hertz}, giving an average efficiency of \SI{97.6}{\percent}.
The experimental Bell value is lower than the quantum prediction because of
noise, most of which is due to unwanted clicks because pulses below the
threshold are close to the threshold, and thus sensitive to small intensity
variations of the lasers.

Adjusting the source to produce fake nonlocal PR boxes~\cite{Popescu1994}
gives a faked Bell value of
\begin{equation}
	S_2=3.6386 \pm 0.0096
  \label{eq:2}
\end{equation}
which is even beyond the quantum bound $2\sqrt2$. The efficiency remains at
\SI{97.6}{\percent}, and noise still lowers the value from the ideal 4.  It
should be noted that Eve is free to combine pulses and phases at will in order
to produce any Bell value between 0 and and the above value. If the noise rate
of the system is known she can compensate by aiming for a higher Bell value,
and letting the noise bring it back down. This allows her to reach a faked
Bell value that is indistinguishable from $2\sqrt2$.

\section{Countermeasures}

Our faked Bell value seemingly violates the Bell-CHSH inequality even though
we are dealing with outcomes from a local realist model. The more appropriate
Bell inequality (\ref{eqn:CHSH_eff}) for conditional correlations is clearly
ineffective as a test of device-independent security with energy-time
entanglement that uses postselection. The bound is too high. We need
to improve the security tests in such a way that they unequivocally show
security, that they can give a loophole-free violation of local realism.

There are two ways to proceed: one is to use fast switching
\cite{Aerts1999,Jogenfors2014}, and the Braunstein-Caves chained Bell
inequalities~\cite{Pearle1970,Braunstein1990} with more terms. The standard chained
inequalities read
\begin{equation}
  \begin{split}
    S_N=&\left|E(A_1B_2)+E(A_3B_2) \right|\\
    &+\left|E(A_3B_4)+E(A_5B_4) \right|\\
    &+\ldots\\
    &+\left|E(A_{2N-1}B_{2N})-E(A_1B_{2N}) \right|\le 2N-2. \label{eqn:BC}
  \end{split}
\end{equation}
In the Franson interferometer with fast switching (F), the chained
inequalities are weakened but still produces a usable bound even after
postselection on coincidence,
\begin{equation}
  \begin{split}
    S_{N,\text{F}}=&\left|E(A_1B_2|\text{coinc.})+E(A_3B_2|\text{coinc.}) \right|\\
    &+\left|E(A_3B_4|\text{coinc.})+E(A_5B_4|\text{coinc.}) \right|\\
    &+\ldots\\
    &+\left|E(A_{2N-1}B_{2N}|\text{coinc.})-E(A_1B_{2N}|\text{coinc.}) \right|\\\le&
    2N-1. \label{eqn:BC-franson}
  \end{split}
\end{equation}
This only gives the upper bound $S_{2,\text{F}}\le3$ for the Bell-CHSH value,
so the standard test is not useful even with fast switching. But the
quantum-mechanical prediction $S_{N,\text{F}}=2N\cos(\pi/{2N})$ does violate
this if $N\geq 3$, even though the violation is smaller than the standard Bell
test. This re-establishes device-independent security for energy-time-entangled
QKD. In practice, though, the requirements are high since the lowest
acceptable visibility is \SI{94.64}{\percent}~\cite{Jogenfors2014}.

A better solution would be to eliminate the core problem: The postselection
loophole.  One alternative is the use of \enquote{hugging}
interferometers~\cite{Cabello2009a} that gives an energy-time-entangled
interferometer with postselection, but without a postselection loophole. The
drawback is the requirement of two fiber links each to Alice and Bob. A Bell
violation has been experimentally shown~\cite{Lima2010}, even with
\SI{1}{\kilo\meter} fiber length~\cite{Cuevas2013}.

\section{Conclusion}
Bell tests are a cornerstone of quantum key distribution and is necessary for
device-independent security. Device-independent Bell inequality violation must
be performed with care to avoid loopholes.  Time-energy-entanglement has the
distinct advantage over polarization that time and energy is more easily
communicated over long distances than polarization. Therefore, time-energy
entanglement may be preferable as quantum resource to perform reliable key
distribution.

In this paper we have shown that quantum key distribution systems based on
energy-time entanglement with postselection are vulnerable to attack if the
corresponding security tests use the original Bell inequality. By blinding the
detectors and using an LHV model Eve lets Alice and Bob think their system
violates Bell's inequality even though she uses classical light. This lets the
attacker fully control the key output and break the security of the system.

Our attack has been performed with a detector efficiency of
\SI{97.6}{\percent} which is high enough to avoid the fair sampling
assumption. We can compare this to Gerhardt et al.~\cite{Gerhardt2011}
where the detector efficiency was \SI{50}{\percent} when
using active basis choice; that attack has an upper limit of
\SI{82.8}{\percent}~\cite{Garg1987,Larsson1998}, while our attack is only
limited to experimental losses. In other words, the attack is possible even with
perfect detection efficiency.

In addition, our attack can produce the unphysical value $S_2=4$ at any
efficiency. It remains a fact that fast switching will restrict this largest
value to 3, but our attack demonstrates the level of control an attacker can
exert onto the system. In order to build a device-independent QKD system based
on energy-time-entanglement the designer will either have to use stronger
tests such as the Braunstein-Caves inequality, or use a system that does not
contain the postselection loophole.

\section*{Acknowledgements}
J.J.\ and J.-\AA.L.\ were supported by CENIIT at LiU.  M.B., J.A., and A.M.E.\
were supported by the Swedish Research Council, Ideas Plus (Polish Ministry of
Science and Higher Education Grant No.\ IdP2011 000361) and ADOPT.

\bibliographystyle{apsrev4-1}
\bibliography{franson}{}

\end{document}